\documentclass[preprint,12pt,authoryear]{elsarticle}

\usepackage{amssymb}
\usepackage{amsmath}

\usepackage[hyphens]{url}
\usepackage{hyperref}
\usepackage{multirow}
\usepackage{makecell}
\usepackage{subcaption} 
\usepackage{color}
\usepackage{xcolor}

\newcommand{\lucas}[1]{\textcolor{black}{#1}}

\journal{Ecological Modelling}

\begin{document}

\begin{frontmatter}



\title{Data-Driven Modelling to predict forest fire spread in the Patagonian region in Argentina} 

\author[balseiro,cab]{Lucas Becerra\corref{cor1}} 
\ead{lucasbecerra0210@gmail.com}
\author[citecca,conicet]{Monica Malen Denham}
\author[balseiro,cab,conicet]{Alejandro B. Kolton}
\author[balseiro,cab,conicet]{Karina Laneri}

\cortext[cor1]{Corresponding author}

\affiliation[balseiro]{organization={Instituto Balseiro. Universidad Nacional de Cuyo},
           city={San Carlos de Bariloche},
           state={Río Negro},
           country={Argentina}}

\affiliation[cab]{organization={Centro Atómico Bariloche. Comisión Nacional de Energía Atómica},
           city={San Carlos de Bariloche},
           state={Río Negro},
           country={Argentina}}

\affiliation[citecca]{organization={Centro Interdisciplinario de Telecomunicaciones, Electrónica, Computación y Ciencia Aplicada (CITECCA). Universidad Nacional de Río Negro},
           city={San Carlos de Bariloche},
           state={Río Negro},
           country={Argentina}}

\affiliation[conicet]{organization={Consejo Nacional de Investigaciones Científicas y Técnicas (CONICET)},
           state={Buenos Aires},
           country={Argentina}}

\begin{abstract}
{Wildfires are among the most severe disturbances affecting forest ecosystems, with over 50,000 hectares burned in Patagonia, Argentina, during 2025 alone. This study implements a Reaction–Diffusion–Convection (RDC) model to simulate wildfire spread in the Steffen and Martín Lakes area, a region severely impacted by fires. By integrating high-resolution maps of slope, wind velocity, and vegetation, we conducted three computational experiments of increasing complexity to simulate fire propagation across heterogeneous landscapes.
\\
We employed a Genetic Algorithm (GA) to recover reference model parameters by maximizing the spatial overlap between simulated and reference burned areas. Subsequently, parameter estimates were refined using XGBoost to improve accuracy. Results demonstrate that the GA accurately recovers reference parameters across all scenarios, while the XGBoost fine-tuning significantly enhances accuracy in simpler cases. This integrated framework offers a systematic approach for estimating difficult-to-measure wildfire parameters, demonstrating the potential of hybrid computational methods for wildfire modeling and forest management.
}
\end{abstract}

\begin{graphicalabstract}
\includegraphics[width=\textwidth]{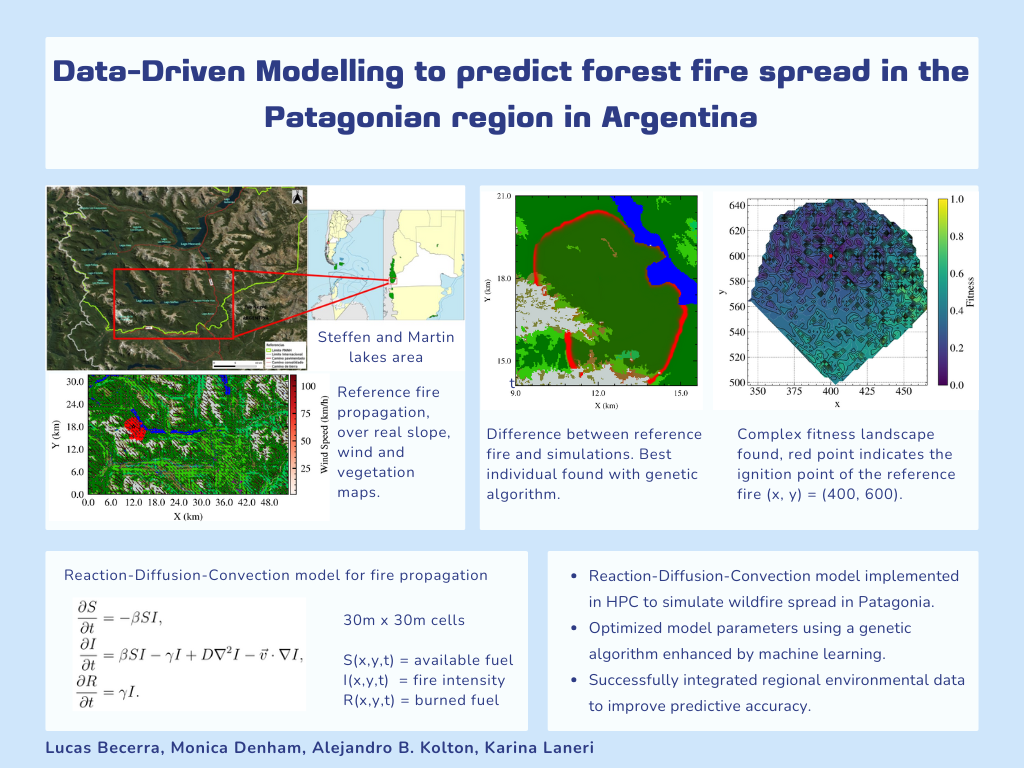}
\end{graphicalabstract}

\begin{highlights}
\item Developed a Reaction-Diffusion-Convection model to simulate wildfire spread in Patagonia.
\item Optimized model parameters using a genetic algorithm enhanced by machine learning.
\item Successfully integrated regional environmental data to improve predictive accuracy.
\end{highlights}


\end{frontmatter}



\section{Introduction}\label{introduction}

Forests represent one of the most important terrestrial ecosystems on the planet. They cover approximately a quarter of the Earth's land area \citep{owid-forest-area} and provide essential functions, including productive, protective, recreative, and environmental functions \citep{fuhrer2000forest}. The loss of forests can lead to the loss of biodiversity, soil erosion by water and wind, contamination of ground and spring water, and even desertification. Despite their crucial roles, forests are increasingly threatened by wildfires, which represent one of the major natural and human-induced disturbances and are expected to intensify with climate change \citep{syphard2015location,kitzberger2022projections}.

Just in the Patagonian Region of Argentina, more than fifty thousand hectares of forest, plantations and residential areas have been lost due to wildfires in 2025, according to official reports \citep{SNMF2025}. 

The drivers behind wildfire ignition are diverse and region-dependent. In Northwestern Patagonia, there is a growing trend towards lightning-caused fires, as recently identified by \cite{kitzberger2025novel}. However, the majority of wildfires are still caused by humans \citep{costafreda2017human}. For instance, $88\%$ of the wildfires that occured between $2016$ and $2020$ in the United States were human-caused \citep{hoover2021wildfire}.   

Anticipating fire behavior is a valuable tool in wildfire management. Traditional approaches such as fire suppression and fuel manipulation are increasingly complemented with alternative strategies, given the growing extenct and intensity of wildfires \citep{syphard2015location}. In that sense, simulations provide a powerful tool to support decision-making. However, in the Patagonian region, wildfire propagation remains a challenge to simulate accurately, since widely used global models (such as \citeauthor{rothermel1972mathematical}, \citeyear{rothermel1972mathematical}) are not adapted to local conditions.

In response to this gap, several efforts have been made to model wildfire propagation in Patagonia. For instance, \cite{morales2015stochastic} proposed a celullar automata model optimized using fire scars from past events in the region. Later, \cite{denham2018using} extended this approach by implementing a parallel genetic algorithm (GA) to recover parameters from synthetic wildfires. A different perspective was adopted by \cite{laneri2020first}, who developed a dynamical model for forest fire behavior in Patagonian landscapes. This model was implemented in a parallel and open-source simulator (Simulador de Incendios Patagonia, \textit{SIP}) to improve the fire managment in the region \citep{denham2022visualization}. Additionally, \cite{barbera2025biotic} contributed to the understanding of the physical drivers of fire through a series of ecological and biophysical studies. Together, these works have significantly expanded our understanding of wildfire dynamics in Patagonia.

\lucas{Beyond regional studies, several works have explored hybrid approaches that combine physics-based wildfire models with machine learning techniques. For instance, \cite{wadhwani2025integrating} and \cite{yu2025probabilistic} implemented surrogate models based on deep neural network architectures to enhance the predictive performance of deterministic and stochastic wildfire models, respectively. Another example is \cite{li2024projecting}, who developed an interptretable hybrid ML framework that explicitly incorporates fuel availability, fuel flammability, and human suppression effects. These approaches aim to improve model predictions when the input parameters are assumed to be known.  In contrast, \cite{jellouli2022impact} employed a cellular automata model coupled with dynamic wind flow simulations, illustrating how rule-based approaches can also capture critical drivers of fire spread without relying on machine learning.}

Nevertheless, parameter optimization remains a critical bottleneck for improving model accuracy, \lucas{ particularly in regions where field measurements are scarce.} In this work, we address this limitation by optimizing a Reaction-Diffussion-Convection (RDC) model for the Patagonian region using a GA-based approach. To this end, we designed three experiments of increasing complexity to test the potential and the limitations of the GA in recovering reference parameters. \lucas{However, the GA is a stochastic optimization method that requires a large number of computationally expensive simulations and does not guarantee convergence to a global optimum, particularly in high-dimensional parameter spaces. To address these limitations, we implemented a surrogate model (XGBoost) to enhance the exploration of the parameter space efficiently, enabling the evaluation of a much larger number of candidate solutions at a significantly lower computational cost.} By combining an RDC model with a GA\lucas{+XGBoost} optimization scheme, this study provides a systematic framework to estimate parameters that are otherwise difficult to measure in the field, contributing to both theoretical understanding and practical fire management in real Patagonian landscapes. 

\lucas{In summary, the novel contributions of this work are: (a) the implementation of a reaction-diffusion-convection (RDC) wildfire model in a realistic Patagonian landscape using spatially explicit data (vegetation, wind, and slope); (b) the development of a controlled experimental framework based on synthetic wildfires to systematically evaluate parameter recovery; (c) a systematic assessment of the performance and limitations of genetic algorithms as the dimensionality and complexity of the parameter space increase; and (d) the integration of a surrogate model (XGBoost) to enhance the exploration of the parameter space.}

\section{Reaction-Diffusion-Convection Model}\label{RDC_model}

The model used in this work is a \emph{reaction--diffusion--convection (RDC)} model \citep{denham2022visualization, laneri2020first}, which describes the spread of fire in a two-dimensional medium in a phenomenological manner. The model does not attempt to capture the microscopic physics of combustion; rather, its purpose is to generate generic reaction--diffusion fronts representative of wildfire propagation at coarse-grained scales. 

The system is described by the following set of equations:
\begin{align}
    \frac{\partial S}{\partial t} &= - \beta S I , \label{eq:susceptible_equation}\\
    \frac{\partial I}{\partial t} &= \beta S I - \gamma I + D \nabla^2 I
    - \vec{v} \cdot \nabla I , \label{eq:infected_equation}\\
    \frac{\partial R}{\partial t} &= \gamma I . \label{eq:recovered_equation}
\end{align}
The fields are defined on a two-dimensional spatial domain.

This spatiotemporal model is inspired by the SIR epidemiological framework
\citep{murray2003mathematical, kolton2019, zagarra2024}. Here, $S$ represents the
local fraction of susceptible vegetation or available fuel, $I$ represents the
fire intensity, and $R$ denotes the fraction of burnt vegetation. The diffusion term $\nabla^2 I$ accounts for the spatial propagation of burning activity, such as heat transfer, flame contact, or ember-induced ignition, within a coarse-grained description, rather than explicit thermal diffusion.

While the mathematical structure of the model is identical to that of spatial SIR epidemic models, its interpretation is fundamentally different. In the present context, \(S\), \(I\), and \(R\) represent local \emph{state fractions} of the medium, rather than conserved populations. Consequently, there is no underlying local conservation law associated with mass or energy in the forest-fire interpretation. A key advantage of Eqs.~(\ref{eq:susceptible_equation})--(\ref{eq:recovered_equation}) over, for instance, cellular automaton models is that they admit analytical results in the simplest homogeneous case, including closed-form expressions for the front velocity and width. Despite its simplicity, it has been shown \cite{zagarra2024} that fronts in this model belong to the same universality class as more realistic models of fire fronts in random media \citep{Provatas1995}, where thermal and chemical processes are taken into account explicitly.

Equation~\eqref{eq:susceptible_equation} describes the dynamics of the susceptible subpopulation. This is governed by a nonlinear reaction term, where $\beta>0$ is the ignition rate and depends on the type of vegetation. A larger $\beta$ means that the vegetation ignites more easily. 

Equation~\eqref{eq:infected_equation} describes the dynamics of fire intensity, associated to burning vegetation. This subpopulation increases at the same rate $\beta$ that susceptibles decrease and burns during a characteristic time $\gamma^{-1}$, transforming into the $R$ fraction. After burning, those cells can not burn again. The spread of fire is further modelled by a diffusion term with a diffusion constant $D$ that for simplicity is assumed homogeneous across the grid. Additionally, an advective term was included into the model, where the advective velocity $\vec{v}$ can be decomposed into the effects of wind $\vec{w}$ and terrain slope $\vec{\nabla} h$:

\begin{equation}
      \vec{v} = A \vec{w} + B \vec{\nabla} h 
\end{equation}

\noindent where $A$ and $B$ are homogeneous constants that weight the contributions of wind and slope of the terrain, respectively. As the fires tend to propagate in the wind direction and in the direction of increasing height \lucas{it is} expected that both, $A>0$ and $B>0$. Finally, Equation~\eqref{eq:recovered_equation} describes the dynamics of the burnt vegetation. The fields $\vec{w}$ and $\vec{\nabla}h$ are input data for the simulations, while the parameters $\beta$, $\gamma$, $D$, $A$, and $B$ are fitted using a genetic algorithm and a fitness function, as will be explained below.

For the numerical implementation, we discretize the set of equations \eqref{eq:susceptible_equation}-\eqref{eq:recovered_equation} on a two-dimensional grid using a finite differences scheme. Reaction and advection terms are integrated explicitly, while diffusion is handled with an alternating direction implicit (ADI) scheme \citep{press2007numerical}. These components are coupled through operator splitting: a half step of the reaction–advection update is followed by a half step of diffusion in one spatial direction, and the process is repeated in the orthogonal direction to complete the time step. Since the advection term is computed explicitly, the advection velocity $\vec{v}$ must follow the Courant-Friedrichs-Lewy (CFL) condition \citep{press2007numerical}. The full discretization is provided in \ref{appendix:numerics}.

\section{Methods}

\subsection{Model Fitting and Synthetic Data}

In order to optimize the RDC model for the Patagonian region, we selected a representative landscape: the area of Steffen and Martin lakes, located at approximately $41^{\circ}30'39.46''$S, $71^{\circ}39'26.34''$W, in Río Negro Province, Argentina; a scenario that has proven to be susceptible to wildfires, with two significant events recorded in December 2021 \citep{bari2022SteffenMartin} and in December 2024 \citep{GobRionegro_Incendio_2025}.

The simulation grid consists of $1748 \times 1060$ square cells, each cell covering an area of $900~\mathrm{m^2}$ ($30~\mathrm{m}$ resolution).
The input data are raster ASCII maps including vegetation, wind, and slope information of the selected landscape. Slope rasters were derived from Digital Elevation Model (DEM) data. The wind direction for each cell was estimated from the average dominant wind direction during fire season and further adjusted for topographic effects using the WindNinja simulator \citep{WindNinja_USFS_2025}. The vegetation map is a raster file with integer values that represent each fuel type. In our simulations, we considered five fuel types: two classes of native forest (Forest A, mainly conifers and broadleaf species, and Forest B, dominated by \textit{Cupressus}), exotic forest, pasture, and shrubland.

We designed three different experiments of increasing complexity in the Steffen and Martin lakes scenario with the objective of testing the GA method to recover the RDC model parameters. In each experiment, using Eqs. \eqref{eq:susceptible_equation}, \eqref{eq:infected_equation} and \eqref{eq:recovered_equation},  we simulated a wildfire 
that will be called ``reference wildfire'', which we generated with a set of arbitrary but plausible parameters, that we aimed to recover using the GA. For each one we performed $500$ simulation steps with a stepsize $dt = 0.5~\mathrm{h}$, which is equivalent to $10.42~\mathrm{days}$ of wildfire propagation. In the following, we describe the three experiments in detail.

\begin{figure}[htbp]
      \centering
      \includegraphics[width=\textwidth]{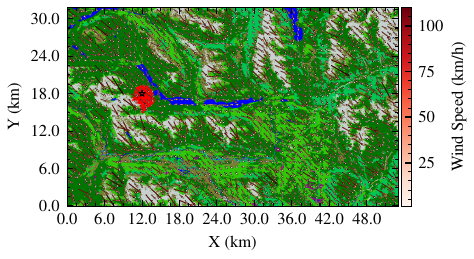}
      \caption{\lucas{Reference wildfire for the first experiment, simulated with $D = 10~\mathrm{m^2~h^{-1}}$, $A = 1 \times 10^{-4}$, $B = 15~\mathrm{m~h^{-1}}$, and ignition point at $(x,y) = (400,600)$. Each green area corresponds to a different fuel type (Forest A, Forest B, exotic forest, pasture, and shrubland), with spatially heterogeneous \(\beta\) and \(\gamma\) values listed in Table~\ref{tab:beta_gamma_values}. Red indicates the final burned area (final state after $500$ steps, $dt = 0.5~\mathrm{h}$), yielding a burnt area of approximately $ 1015~\mathrm{ha}$; the star marks the ignition point; arrows indicate wind direction.
      }}
      \label{fig:first_experiment_reference_fire}
\end{figure}

In the first experiment, we optimized the parameters $D$, $A$, $B$, and the ignition point coordinates $(x,y)$, while keeping $\beta$ and $\gamma$ fixed but heterogeneous in space according to the vegetation map. The values of $\beta$ and $\gamma$ for each fuel type are listed in Table \ref{tab:beta_gamma_values}. The reference wildfire was simulated with the parameters $D = 10~\mathrm{m^2~h^{-1}}$, $A = 1 \times 10^{-4}$, $B = 15~\mathrm{m~h^{-1}}$, and ignition point at $(x,y) = (400,600)$. The reference wildfire for this experiment is shown in Figure \ref{fig:first_experiment_reference_fire}. In total, we recovered five parameters $(D,A,B,x,y)$ in this experiment.

\begin{table}[htbp]
      \centering
      \begin{tabular}{c|c|c}
            Fuel type & $\beta$ value $(\mathrm{h^{-1}})$ & $\gamma$ value $(\mathrm{h^{-1}})$ \\ \hline
            Forest A & $0.91$ & $0.50$ \\ 
            Forest B & $0.72$ & $0.38$ \\
            Exotic forest & $1.38$ & $0.84$ \\
            Pasture & $1.94$ & $0.45$  \\
            Shrubland & $0.75$ & $0.14$  \\
      \end{tabular}
      \caption{Values for the ignition rate $\beta$ and the extinction rate $\gamma$ for each fuel type considered in the first and third experiment.
      }
      \label{tab:beta_gamma_values}
\end{table}

\begin{figure}[htbp]
      \centering
      \includegraphics[width=\textwidth]{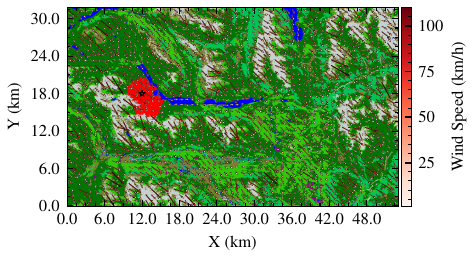}
      \caption{\lucas{Reference wildfire for the second experiment (Steffen-Martin landscape). The same reference values as the first experiment were used for ($D = 10~\mathrm{m^2~h^{-1}}$, $A = 1 \times 10^{-4}$, $B = 15~\mathrm{m~h^{-1}}$, and ignition point at $(x,y) = (400,600)$), while homogeneous $\beta$ and $\gamma$ were introduced as additional tunable parameters with reference values ($\beta = 1.5~\mathrm{h^{-1}}$ and $\gamma = 0.5~\mathrm{h^{-1}}$), respectively. Red indicates burned area (final state after $500$ steps, $dt = 0.5~\mathrm{h}$), yielding a burnt area of approximately $2342~\mathrm{ha}$.}}
      \label{fig:second_experiment_reference_fire}
\end{figure}

In the second experiment, we optimized the same parameters (with the same values) as in the first one, but also including $\beta$ and $\gamma$ as free parameters. For simplicity, the landscape was considered homogeneous, meaning that the same ignition and extinction rates were applied to all fuel types. The reference values used were $\beta = 1.5~\mathrm{h^{-1}}$ and $\gamma = 0.5~\mathrm{h^{-1}}$. The resulting reference wildfire is shown in Figure \ref{fig:second_experiment_reference_fire}. In total, we recovered seven parameters $(D,A,B,x,y,\beta,\gamma)$ in this experiment.

\begin{figure}[htbp]
      \centering
      \includegraphics[width=\textwidth]{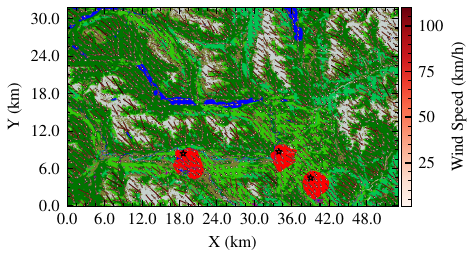}
      \caption{\lucas{Reference wildfire used in Experiment 3 (Steffen-Martin landscape). The simulation uses $D=10~\mathrm{m^2~h^{-1}}$, $A=1\times10^{-4}$, and $B=15~\mathrm{m~h^{-1}}$, with three fixed ignition points at $(1130,290)$, $(1300,150)$, and $(620,280)$. Each green area corresponds to a different fuel type (Forest A, Forest B, exotic forest, pasture, and shrubland), with heterogeneous and tunable $\beta_i$ and $\gamma_i$ values (Table~\ref{tab:beta_gamma_values}). Red indicates the final burned area (final state after $500$ steps, $dt = 0.5~\mathrm{h}$), yielding a burnt area of approximately $4241~\mathrm{ha}$; this reference fire was designed to burn a representative fraction of all fuel types.}}
      \label{fig:third_experiment_reference_fire}
\end{figure}

In the third and more complex experiment, we added a different ignition and extinction rate for each fuel type as free parameters. The reference values are reutilized from Table \ref{tab:beta_gamma_values}. In order to optimize the five fuel types, we had to design a proper wildfire, meaning that the burnt area had to cover a representative area of each fuel type to be sensible to changes in the free parameters. To accomplish that challenge, we designed a wildfire with multiple ignition points so it could cover a representative area of each fuel type. To avoid an increase in the number of free parameters, we fixed the coordinates of the ignition points. For this case, we simulated a wildfire with three ignition points in the cells $(1130, 290)$, $(1300,150)$, $(620,280)$. The other parameters are the same as those in the first and second experiments. The resulting reference wildfire is shown in Figure \ref{fig:third_experiment_reference_fire}. Its burnt area covered $29.15\%$ of forest A, $44.09\%$ of forest B, $3.62\%$ of exotic forest, $13.18\%$ of pasture and $9.96\%$ of shrubland, ensuring a representative impact across all fuel types. In total we recovered thirteen parameters $(D,A,B,\beta_i,\gamma_i) \; \text{for} \; i = 1,2,3,4,5$ in this experiment.

The range of parameters to generate the initial population for the GA was set following different criteria. For the advection constants $A$ and $B$, we took into account the maximum possible value given by the CFL condition. For a half stepsize of $dt/2 = 0.25~\mathrm{h}$ (the ADI scheme requires an evaluation at half time steps), a cell size of $30~\mathrm{m}$ and taking the maximum value of wind and slope, the maximum values are $A_{\mathrm{max}}=5.918 \times 10^{-4}$ and $B_{\mathrm{max}} = 24.853~\mathrm{m~h^{-1}}$. For the diffusion constant, we set a maximum value of $D_{\mathrm{max}} = 100~\mathrm{m^2~h^{-1}}$. For the $\beta$ and $\gamma$ rates, we set a maximum value of $2~\mathrm{h^{-1}}$ and $1~\mathrm{h^{-1}}$, respectively. All these values were generated uniformly between zero and the maximum value. The ignition coordinates (in the first and second case) were limited to the burnt area of the reference wildfire.

As a summary, the details of the tree experiments are listed in Table \ref{tab:summary_synthetic_experiments}.

\begin{table}[htbp]
\centering
\begin{tabular}{c||c|c||c|c}
\textbf{Exp.} & \textbf{\begin{tabular}[c]{@{}l@{}}Parameter \\ to adjust\end{tabular}} & \textbf{\begin{tabular}[c]{@{}l@{}}Reference\\ value\end{tabular}} & \textbf{\begin{tabular}[c]{@{}l@{}}Fixed \\ parameters\end{tabular}} & \textbf{\begin{tabular}[c]{@{}l@{}}Reference\\ value\end{tabular}} \\ \hline
\multirow{4}{*}{1} & $D$ & $10~\mathrm{m^{2}~h^{-1}}$ & $\beta$ & \begin{tabular}[c]{@{}l@{}}Heterogeneous\\  according to \\ Table \ref{tab:beta_gamma_values}\end{tabular} \\ 
 & $A$ & $1 \times 10^{-4}$ & $\gamma$ & \begin{tabular}[c]{@{}l@{}}Heterogeneous \\ according to \\ Table \ref{tab:beta_gamma_values}\end{tabular} \\ 
 & $B$ & $15~\mathrm{m~h^{-1}}$ & - & - \\  
 & $(x,y)$ & $(400,600)$ & - & - \\ \hline
\multirow{6}{*}{2} & $D$ & $10~\mathrm{m^{2}~h^{-1}}$ & - & - \\ 
 & $A$ & $1 \times 10^{-4}$ & - & - \\ 
 & $B$ & $15~\mathrm{m~h^{-1}}$ & - & - \\ 
 & $(x,y)$ & $(400,600)$ & - & - \\ 
 & $\beta$ & $1.5~\mathrm{h^{-1}}$ & - & - \\ 
 & $\gamma$ & $0.5~\mathrm{h^{-1}}$ & - & - \\ \hline
\multirow{5}{*}{3} & $D$ & $10~\mathrm{m^{2}~h^{-1}}$ & $(x_1,y_1)$ & $(1130,290)$ \\ 
 & $A$ & $1 \times 10^{-4}$ & $(x_2,y_2)$ & $(1300,150)$ \\  
 & $B$ & $15~\mathrm{m~h^{-1}}$ & $(x_3,y_3)$ & $(620,280)$ \\  
 & $\beta$ & \begin{tabular}[c]{@{}l@{}}Heterogeneous \\ according to \\ Table \ref{tab:beta_gamma_values}\end{tabular} & - & - \\ 
 & $\gamma$ & \begin{tabular}[c]{@{}l@{}}Heterogeneous \\ according to \\ Table \ref{tab:beta_gamma_values}\end{tabular} & - & - \\ 
\end{tabular}
\caption{Summary of the three synthetic experiments designed to evaluate the proposed fitting methods. The parameters to be adjusted and their reference values are detailed, as well as the fixed parameters and their respective values.}
\label{tab:summary_synthetic_experiments}
\end{table}

\subsection{Genetic Algorithm}

To estimate the optimal parameters of the RDC model, we implemented a GA. This population-based stochastic optimization method is well suited for high-dimensional, non-linear problems in which the objective function is computationally expensive to evaluate \citep{katoch2021review}.

The algorithm begins by generating an initial population of individuals, where each individual is defined by a set of parameters (genes). Every individual is then evaluated through a fitness function, which quantifies the similarity between the simulated fire spread and a reference map. In this work, we adopted the fitness function proposed in \citet{denham2009prediccion}, which compares the number of burnt cells in common between the two maps:

\begin{equation} \label{eq:fitness_function}
      \delta = \frac{A \cup A^{*} - A \cap A^{*}}{A} \, ,
\end{equation}

\noindent where $A$ denotes the set of burnt cells in the reference map and $A^{*}$ the set of burnt cells in the simulated map. Equation~\eqref{eq:fitness_function} approaches zero as the simulated map becomes more similar to the reference. In the worst case in which the simulated map has a null intersection with the reference, $\delta=(A+A^*)/A = 1+A^*/A>1$ and the maximum value is given by the size of the map. We considered a burnt cell any cell with $R > 0.001$ at the end of the simulation. 

Once the fitness of all individuals is computed, the GA applies three main operators: selection, where the best candidates are chosen according to their fitness; crossover, where pairs of individuals exchange some of their genes with a certain probability; and mutation, which introduces small random changes to the genes. These steps are repeated over successive generations, progressively improving the population. The population is constant in size throughout the generations.

Further implementation details of the GA are provided in \ref{appendix:ga}.

\subsection{Parameters search refinement and error assessment}

Since the GA requires a large number of \lucas{computationally expensive} wildfire simulations to explore the parameter space, \lucas{and does not guarantee convergence to a global optimum}, we implemented a refinement method to approximate the fitness function and find parameter combinations with lower fitness values. This approach takes advantage of the million simulations performed in the three experiments to train a machine learning model that can predict the fitness of new parameter sets without running the full simulation. \lucas{Among avalaible surrogate modeling approaches, XGBoost stands out for its ability to handle large datasets and complex relationships \citep{chen2016xgboost,grinsztajn2022tree,shwartz2022tabular}.}

We trained an XGBoost regressor \citep{chen2016xgboost} using the parameters as input features and the corresponding fitness values as targets. The dataset was split into training ($80\%$) and testing ($20\%$) sets. The model's performance was evaluated using the mean squared error (MSE) and the coefficient of determination ($R^2$). The trained model used $5000$ estimators, a learning rate of $0.01$, a maximum depth of $10$, L1 regularization term of $0.1$, and L2 regularization term of $1$. \lucas{This hyperparameter configuration was selected based on a grid search procedure, and applied consistently across the three experiments. Preliminary tuning efforts, particularly for the third experiment, did not lead to significant performance improvements.} 

With the trained surrogate model, we can rapidly evaluate the fitness of new parameter sets, enabling a more efficient exploration of the parameter space. We evaluated $100$ million candidates using the surrogate model for each experiment. Then, we selected the top $50$ candidates with the lowest predicted fitness and performed full wildfire simulations to compute their true fitness values. \lucas{The accuracy of this approach depends on the surrogate model’s ability to approximate the fitness landscape, which may degrade in highly complex or high-dimensional cases.}

To quantify the uncertainty in the parameters estimated by the GA and the XGBoost model, we applied a bootstrap analysis to the final population of each experiment. This resampling technique allows us to approximate the sampling distribution of the minimum fitness values, and consequently the distribution of the best-fit parameters. For each bootstrap replicate, we sampled with replacement from the final population, extracted the best individual (minimum fitness), and repeated the process $1000$ times. From these replicates we computed the mean and the $95\%$ confidence interval, defined by the $2.5$ and $97.5$ percentiles.

\lucas{The estimated uncertainty reflects the variability in the best-fit parameters arising from the stochastic nature of the genetic algorithm and the structure of the fitness landscape. In this context, the bootstrap confidence intervals (defined by the $2.5$ and $97.5$ percentiles of the estimated distribution) provide an empirical measure of the range of parameter values that yield similarly optimal solutions. A full propagation of parameter uncertainty into model outputs is beyond the scope of this work.}

\section{Results}

\begin{figure}[htbp]
    \centering
    \includegraphics[width=0.7\textwidth]{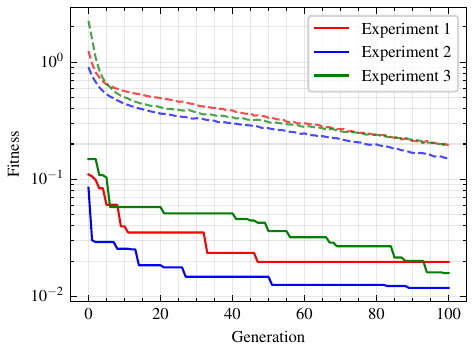}
    \caption{Evolution of the fitness in the three experiments. The dashed line shows the mean fitness per generation, while the solid line indicates the minimum fitness (best individual). Elitism ensures that the best individual is preserved across generations.}
    \label{fig:fitness_evolution}
\end{figure}

We ran the genetic algorithm with $100$ generations and a population size of $10000$ individuals for each experiment, resulting in $1 \times 10^6$ wildfire simulations per experiment. Figure \ref{fig:fitness_evolution} shows the evolution of the fitness along generations. The dashed line corresponds to the population mean, while the solid line represents the minimum fitness, i.e. the best individual of each generation. Due to elitism, the best individual is always preserved, which guarantees a non-increasing trend of the minimum fitness.

The bootstrap results for the three experiments are reported in Table \ref{tab:bootstrap_genetic_algorithm}. Each table lists the optimized parameters, their bootstrap mean, the corresponding $95\%$ confidence intervals, the relative uncertainty, defined as the half-width of the confidence interval divided by the mean, \lucas{and the relative error with respect to the true values}.

\begin{table}[htbp]
\centering
\small
\begin{tabular}{c|c|c|c|c|c}
\textbf{Exp.} &
\textbf{Parameter} &
\textbf{\makecell{Bootstrap \\ Mean}} &
\textbf{\makecell{Confidence \\ Interval}} &
\textbf{\makecell{Uncertainty \\ (\%)}} &
\textbf{\makecell{Relative \\ Error (\%)}} \\ \hline

\multirow{5}{*}{1}
& $D~(\mathrm{m^2~h^{-1}})$ & $11.221$ & $(10.436,\;12.827)$ & $10.65$ & $12.21$ \\
& $A$ & $1.209\times 10^{-4}$ & $(0.934,\;1.712)\times 10^{-4}$ & $32.18$ & $20.90$ \\
& $B~(\mathrm{m~h^{-1}})$ & $17.367$ & $(14.345,\;22.786)$ & $24.30$ & $15.78$ \\
& $x$ & $398.576$ & $(396.000,\;400.000)$ & $0.50$ & $0.36$ \\
& $y$ & $602.848$ & $(600.000,\;608.000)$ & $0.66$ & $0.47$ \\ \hline

\multirow{7}{*}{2}
& $D~(\mathrm{m^2~h^{-1}})$ & $34.056$ & $(30.278,\;34.670)$ & $6.45$ & $240.56$ \\
& $A$ & $1.107\times 10^{-4}$ & $(1.069,\;1.121)\times 10^{-4}$ & $2.35$ & $10.70$ \\
& $B~(\mathrm{m~h^{-1}})$ & $17.955$ & $(14.652,\;18.511)$ & $10.75$ & $19.70$ \\
& $x$ & $402.784$ & $(402.000,\;404.000)$ & $0.25$ & $0.70$ \\
& $y$ & $600.423$ & $(600.000,\;604.000)$ & $0.17$ & $0.07$ \\
& $\beta~(\mathrm{h^{-1}})$ & $1.002$ & $(0.990,\;1.088)$ & $4.89$ & $33.20$ \\
& $\gamma~(\mathrm{h^{-1}})$ & $0.418$ & $(0.411,\;0.480)$ & $8.25$ & $16.4$ \\ \hline

\multirow{15}{*}{3}
& $D~(\mathrm{m^2~h^{-1}})$ & $11.321$ & $(9.884,\;11.899)$ & $8.91$ & $13.21$ \\
& $A$ & $1.008\times 10^{-4}$ & $(0.953,\;1.036)\times 10^{-4}$ & $4.13$ & $0.80$ \\
& $B~(\mathrm{m~h^{-1}})$ & $14.527$ & $(13.211,\;15.075)$ & $6.43$ & $3.15$ \\
& $\beta_1~(\mathrm{h^{-1}})$ & $0.527$ & $(0.503,\;0.621)$ & $11.18$ & $42.09$ \\
& $\beta_2~(\mathrm{h^{-1}})$ & $0.587$ & $(0.538,\;0.609)$ & $6.04$ & $18.47$ \\
& $\beta_3~(\mathrm{h^{-1}})$ & $0.898$ & $(0.764,\;1.068)$ & $16.97$ & $34.93$ \\
& $\beta_4~(\mathrm{h^{-1}})$ & $1.969$ & $(1.955,\;2.000)$ & $1.13$ & $1.49$ \\
& $\beta_5~(\mathrm{h^{-1}})$ & $0.658$ & $(0.587,\;0.771)$ & $13.94$ & $12.27$ \\
& $\gamma_1~(\mathrm{h^{-1}})$ & $0.146$ & $(0.129,\;0.241)$ & $38.33$ & $70.80$ \\
& $\gamma_2~(\mathrm{h^{-1}})$ & $0.259$ & $(0.168,\;0.286)$ & $22.92$ & $31.84$ \\
& $\gamma_3~(\mathrm{h^{-1}})$ & $0.363$ & $(0.209,\;0.602)$ & $54.61$ & $56.79$ \\
& $\gamma_4~(\mathrm{h^{-1}})$ & $0.530$ & $(0.471,\;0.559)$ & $8.31$ & $15.78$ \\
& $\gamma_5~(\mathrm{h^{-1}})$ & $0.113$ & $(0.100,\;0.151)$ & $22.66$ & $84.93$ \\
\end{tabular}
\caption{Results of the bootstrap analysis of the genetic algorithm for the three synthetic experiments. The uncertainty corresponds to the relative half-width of the confidence interval, calculated as $(\text{Upper Limit} - \text{Lower Limit}) / (2 \times \text{Mean}) \times 100$. The reference values for each experiment are found in Table \ref{tab:summary_synthetic_experiments}.}
\label{tab:bootstrap_genetic_algorithm}
\end{table}

The results reported in the Tables \ref{tab:bootstrap_genetic_algorithm} show that the GA was able to recover some reference parameters with good accuracy across the three experiments. 

After running the GA, we trained the XGBoost model for each experiment \lucas{using the parameters as input features and the fitness values obtained from the GA million simulations as targets}. The performance of the trained model for the three experiments is summarized in Table \ref{tab:surrogate_model_performance} and the results of the bootstrap analysis for the best individuals obtained after the refinement procedure are reported in Table \ref{tab:bootstrap_xgboost_results}.

\begin{table}
      \centering
      \begin{tabular}{c|c|c}
            Experiment & MSE & $R^2$ \\ \hline
            First & $(1.14 \pm 0.02) \times 10^{-4}$ & $0.99746 \pm 0.00004$ \\ 
            Second & $(8.19 \pm 0.07) \times 10^{-4}$ & $0.98084 \pm 0.00013$ \\ 
            Third & $(236 \pm 4) \times 10^{-4}$ & $0.872 \pm 0.003$ \\ 
      \end{tabular}
      \caption{Performance metrics of the refined search for the three experiments. A cross-validation with $5$ folds was performed to compute the mean and standard deviation of the metrics.}
      \label{tab:surrogate_model_performance}
\end{table}

\begin{table}[htbp]
\centering
\small
\begin{tabular}{c|c|c|c|c|c}
\textbf{Exp.} &
\textbf{Parameter} &
\textbf{\makecell{Bootstrap \\ Mean}} &
\textbf{\makecell{Confidence \\ Interval}} &
\textbf{\makecell{Uncertainty \\ (\%)}} &
\textbf{\makecell{Relative \\ Error (\%)}} \\ \hline

\multirow{5}{*}{1}
& $D~(\mathrm{m^2~h^{-1}})$ & $10.206$ & $(9.846,\;10.361)$ & $2.52$ & $2.06$ \\
& $A$ & $1.008\times 10^{-4}$ & $(0.884,\;1.352)\times 10^{-4}$ & $25.81$ & $0.80$ \\
& $B~(\mathrm{m~h^{-1}})$ & $13.866$ & $(12.622,\;18.910)$ & $22.78$ & $7.56$ \\
& $x$ & $398.930$ & $(396.185,\;399.465)$ & $0.41$ & $0.27$ \\
& $y$ & $600.353$ & $(599.637,\;603.788)$ & $0.34$ & $0.06$ \\ \hline

\multirow{7}{*}{2}
& $D~(\mathrm{m^2~h^{-1}})$ & $22.772$ & $(20.286,\;25.631)$ & $11.74$ & $127.72$ \\
& $A$ & $1.522\times 10^{-4}$ & $(1.038,\;1.615)\times 10^{-4}$ & $18.95$ & $52.2$ \\
& $B~(\mathrm{m~h^{-1}})$ & $19.873$ & $(15.646,\;20.478)$ & $12.16$ & $32.49$ \\
& $x$ & $397.056$ & $(394.909,\;399.363)$ & $0.56$ & $0.74$ \\
& $y$ & $607.966$ & $(597.516,\;610.901)$ & $1.10$ & $1.33$ \\
& $\beta~(\mathrm{h^{-1}})$ & $1.040$ & $(0.838,\;1.128)$ & $13.96$ & $30.67$ \\
& $\gamma~(\mathrm{h^{-1}})$ & $0.337$ & $(0.125,\;0.424)$ & $44.50$ & $32.60$ \\ \hline

\multirow{15}{*}{3}
& $D~(\mathrm{m^2~h^{-1}})$ & $48.113$ & $(37.493,\;96.937)$ & $61.78$ & $381.13$ \\
& $A$ & $1.415\times 10^{-4}$ & $(1.178,\;1.895)\times 10^{-4}$ & $25.33$ & $41.50$ \\
& $B~(\mathrm{m~h^{-1}})$ & $18.833$ & $(8.024,\;23.183)$ & $40.25$ & $25.55$ \\
& $\beta_1~(\mathrm{h^{-1}})$ & $0.823$ & $(0.680,\;0.918)$ & $14.50$ & $9.56$ \\
& $\beta_2~(\mathrm{h^{-1}})$ & $0.935$ & $(0.862,\;0.966)$ & $5.53$ & $29.86$ \\
& $\beta_3~(\mathrm{h^{-1}})$ & $0.737$ & $(0.267,\;0.904)$ & $43.20$ & $46.59$ \\
& $\beta_4~(\mathrm{h^{-1}})$ & $1.310$ & $(0.881,\;1.467)$ & $22.35$ & $32.47$ \\
& $\beta_5~(\mathrm{h^{-1}})$ & $0.206$ & $(0.136,\;0.476)$ & $82.43$ & $72.53$ \\
& $\gamma_1~(\mathrm{h^{-1}})$ & $0.725$ & $(0.541,\;0.791)$ & $17.27$ & $45.00$ \\
& $\gamma_2~(\mathrm{h^{-1}})$ & $0.731$ & $(0.687,\;0.826)$ & $9.49$ & $92.37$ \\
& $\gamma_3~(\mathrm{h^{-1}})$ & $0.707$ & $(0.334,\;0.849)$ & $36.33$ & $15.83$ \\
& $\gamma_4~(\mathrm{h^{-1}})$ & $0.470$ & $(0.094,\;0.834)$ & $78.68$ & $4.44$ \\
& $\gamma_5~(\mathrm{h^{-1}})$ & $0.735$ & $(0.683,\;0.871)$ & $12.78$ & $425.00$ \\
\end{tabular}
\caption{Bootstrap analysis results for candidates found by the XGBoost model and validated with the RDC model for the three synthetic experiments. Reference values for each experiment are found in Table \ref{tab:summary_synthetic_experiments}.}
\label{tab:bootstrap_xgboost_results}
\end{table}

\lucas{In order to quantify the differences between the GA optimization and the XGBoost refinement, we compared the mean fitness values obtained from both methods in Table \ref{tab:fitness_comparison}, using a bootstrap analysis. The reported mean fitness corresponds to the expected value of the best fitness under the empirical sampling distribution obtained through bootstrap resampling, providing a robust estimate of the typical performance of each method. In contrast to reporting a single minimum fitness value, this approach reduces sensitivity to stochastic fluctuations and outliers in the optimization process.}

\begin{table}[htbp]
\small
\centering
\begin{tabular}{c|c|c} 
Experiment & Genetic Algorithm & XGBoost \\ \hline
$1$ & $0.028$ & $0.023$ \\ \hline
$2$ & $0.012$ & $0.022$ \\ \hline
$3$ & $0.020$ & $0.398$ \\
\end{tabular}
\caption{\lucas{Comparison of mean fitness values obtained from the GA and XGBoost methods across the three synthetic experiments, based on bootstrap analysis. The XGBoost method outperforms the GA in the first experiment, while the GA achieves better performance in the second and third experiments.}}
\label{tab:fitness_comparison}
\end{table}

\lucas{The results in Table \ref{tab:fitness_comparison} show that the XGBoost refinement improves upon the GA results in the first experiment. In contrast, the GA achieves better performance in the second experiment, and significantly outperforms XGBoost in the third one. This behavior reflects the increasing complexity of the fitness landscape: while the surrogate model is effective in simpler scenarios, its performance deteriorates in more complex and high-dimensional cases, where accurately approximating the fitness function becomes more challenging.}

\section{Discussion}

In this section, we analyze the performance of the genetic algorithm and the subsequent refinement method across the three experiments, focusing on which parameters were more reliably estimated and which ones showed larger uncertainties. By contrasting these outcomes, we highlight the strengths and limitations of the method in recovering the true model parameters.

In the first and second experiment, the ignition point coordinates were the best-estimated parameters. This outcome can be attributed to the fact that the search space for the coordinates was discrete and restricted to the burnt area of the reference wildfire. In the second experiment, however, the bootstrap mean for the diffusion constant exhibited a relative error larger than $90\%$, highlighting a difficulty of optimizing this parameter. Moreover, the inclusion of $\beta$ and $\gamma$ as additional tunable parameters further increased the complexity of the fitness function, making convergence more challenging.

\begin{figure}[htbp]
      \centering
      \includegraphics[width=0.5\textwidth]{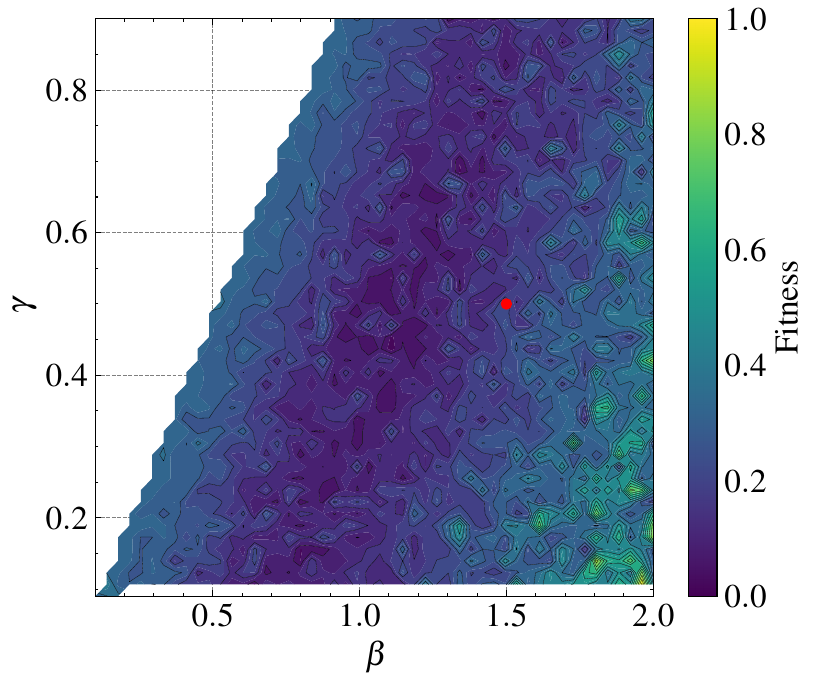}
      \caption{Fitness landscape in relation to $\beta$ and $\gamma$ for the second case. It is shown that there is a valley of similar fitness values, which creates a \lucas{strong} correlation between both parameters and hinders their accurate estimation. The red point indicates the reference values $(\beta,\gamma)=(1.5,0.5)$.}
      \label{fig:surf_beta_gamma_exp2}
\end{figure}

\begin{figure}[htbp]
      \centering
      \includegraphics[width=0.5\textwidth]{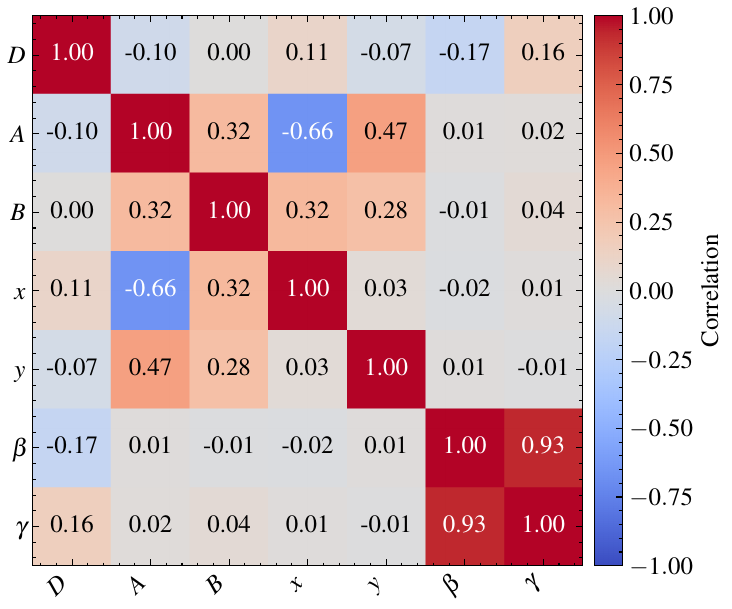}
      \caption{\lucas{Correlation matrix of the model parameters found in the second experiment. The correlation was calculated using the best $10000$ individuals across all the generations. There is a strong correlation between $\beta$ and $\gamma$, which leads to almost a linear relation between both parameters and difficults the optimization process.}}
      \label{fig:correlation_matrix_exp2}
\end{figure}

Figure \ref{fig:surf_beta_gamma_exp2} illustrates the fitness landscape with respect to $\beta$ and $\gamma$ in the second experiment. The surface exhibits a highly irregular structure, with many points of similar fitness and no clear decreasing direction, which makes accurate estimation difficult. \lucas{This behavior is consistent with the correlation matrix for the second experiment (Fig. \ref{fig:correlation_matrix_exp2}), where $\beta$ and $\gamma$ show a strong correlation ($0.93$), indicating an almost linear relationship between them. This coupling can be attributed to the role of the basic reproduction rate in diffusive and non-diffusive epidemic models \citep{murray2003mathematical}.}

\lucas{The correlation analysis supports the presence of practical non-identifiability of $\beta$ and $\gamma$ in the second experiment: many parameter combinations yield similarly low fitness, forming an elongated valley in the objective landscape. This indicates that the available data are insufficient to uniquely constrain these parameters. Thus, the main limitation in our current setup is practical identifiability, although additional structural analysis would be required to formally rule out model intrinsic non-identifiability.}

\begin{figure}[htbp]
      \centering
      \begin{subfigure}{0.45\textwidth}
            \centering
            \includegraphics[width=\textwidth]{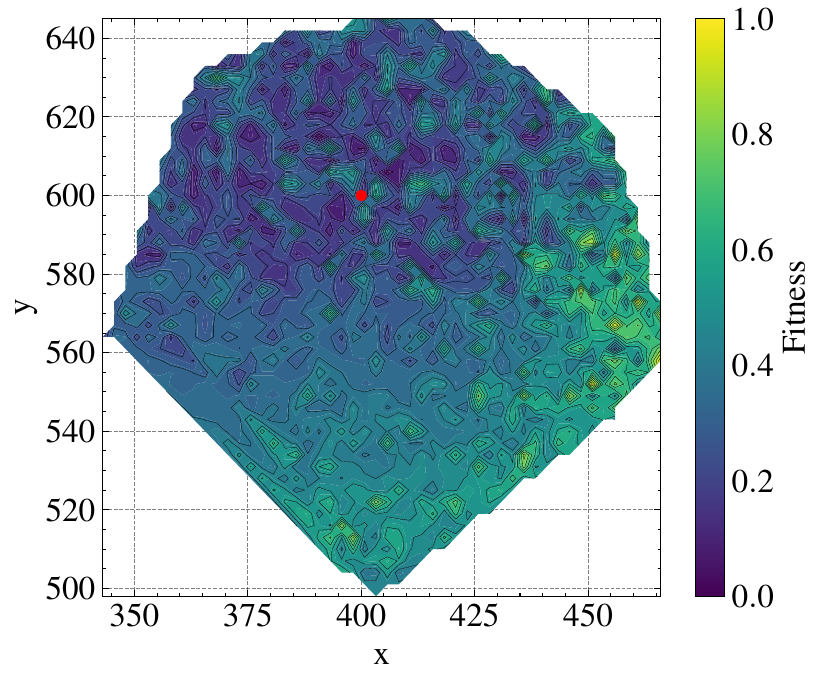}
      \end{subfigure}
      \hfill
      \begin{subfigure}{0.45\textwidth}
            \centering
            \includegraphics[width=\textwidth]{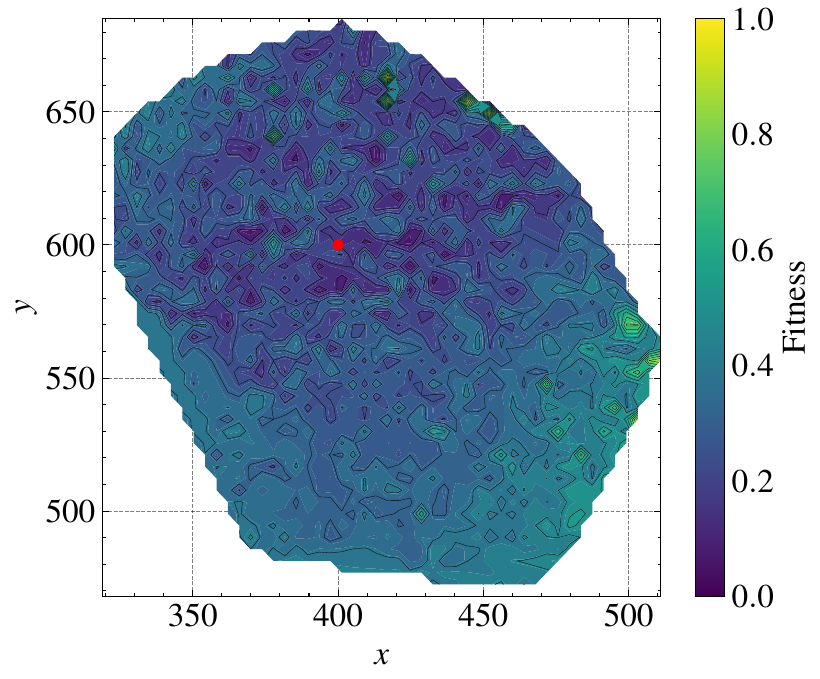}
      \end{subfigure}
      \caption{Fitness landscape in relation to the first (left) and second (right) experiment. In contrast with Figure \ref{fig:surf_beta_gamma_exp2}, there is no ambiguity in the estimation of the ignition coordinates. The red point indicates the reference values $(x,y)=(400,600)$.}
      \label{fig:surf_x_y_exps_1_2}
\end{figure}

By contrast, Figure \ref{fig:surf_x_y_exps_1_2} shows the fitness landscape for the ignition coordinates in the first and second experiments. In both cases, in contrast to the previous figure, there is no ambiguity in the estimation of the coordinates. Despite the irregularity of the fitness landscape, the GA was able to converge to the correct ignition point.

\begin{figure}[htbp]
      \centering
      \includegraphics[width=0.5\textwidth]{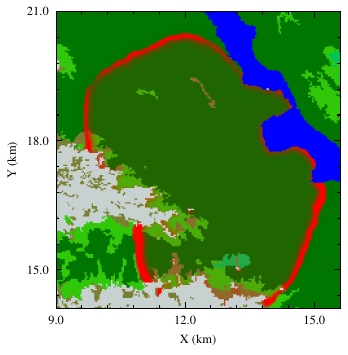}
      \caption{Difference between the reference wildfire and the best-estimated wildfire in the second experiment. The difference is seen in the border of the burnt area, which indicates that both fire spread patterns are very similar.}
      \label{fig:second_difference_map}
\end{figure}

Despite the fact that the diffusion constant was poorly estimated in the second experiment, the fitness of the best individual was very close to that of the reference wildfire, as can be seen in Figure \ref{fig:fitness_evolution}. This result suggests that different combinations of parameters can lead to similar fire spread patterns, at least in this regime of the parameter space. The difference between the reference and the best-estimated parameters for the second experiment is shown in Figure \ref{fig:second_difference_map}.

In the third experiment, the diffusion and advection constants were accurately estimated. Not only that, but the associated relative errors were smaller than in the previous experiments. Nevertheless, the ignition and extinction rates for the different fuel types were poorly estimated in several cases. This result can be partly explained by the vegetation distribution in the landscape. Surprisingly, the coefficient for pasture was the best estimated, even though these were not the most prominent vegetation within the burnt area. Moreover, the outcome of a single GA run may lead to a misleading interpretation; however, this limitation was mitigated through the bootstrapping analysis. 

The implementation of the refinement method using machine learning demonstrated promising results, particularly in the first experiment, where better estimations of the parameters were achieved compared to the GA without refinement. 
This outcome can be attributed to the high accuracy of the refined method in approximating the fitness function, as indicated by the $R^2$ value of $0.99746$. In the second experiment, the results were comparable to those obtained with the GA, as the fitness landscape conveged more clearly to a minimum. It is shown that the GA converged to a close minimum, and the XGBoost model was able to identify similar candidates. Then, the bias in the GA was transferred to the XGBoost model. In the third experiment, however, the refinement method produced worse results than the GA ($R^2$ = $0.872$). The increased complexity of the fitness landscape in the third experiment likely contributed to this reduced accuracy, as can be seen in Table \ref{tab:surrogate_model_performance}. Further investigation is needed to enhance the refinement method for high-dimensional parameter spaces with complex fitness landscapes.

\lucas{From an ecological perspective, the parameters of the RDC model have a direct interpretation in terms of wildfire behavior. For instance, the parameter $\beta$ is associated with the efective spread rate of the fire, while $\gamma$ represents the rate of fuel depletion or fire extinction. The diffusion term captures the local spread due to short-range processes such as radiation and spotting, whereas the convection term accounts for the influence of wind and slope in fire propagation.}

\lucas{In this work, we used a fitness function (Eq. \ref{eq:fitness_function}) based on the spatial overlap between the final burned areas of the simulated and reference fire spread patterns. In this case, the reference corresponds to a synthetic wildfire generated by the SIR model. Alternative approaches in the literature incorporate temporal information into the fitness function \citep{kelso2015techniques}, which can capture aspects of the fire dynamics beyond the final burned extent. This is ecologically relevant because two wildfires may produce similar final burn scars while exhibiting different temporal evolutions, such as differences in front velocity or rate spread \citep{kolaitis2023comparative}.}

\lucas{In the model used in this work, however, time has a direct physical interpretation, as the system is based on a physical reaction-diffusion-convection formulation. Consequently, simulations are run for the same duration as the reference fire, which implicitly constrains the average propagation speed of the fire front. This partially reduces the space of admissible solutions, as similar burned areas obtained at the same simulation time cannot correspond to arbitrarily different average propagation velocities. However, this costraint is not sufficient to uniquely determine the underlying parameters, since different combinations can still produce similar spatial patterns and comparable effective propagation regimes. Differences in the detailed temporal dynamics of the fire spread may still exist. Exploring fitness functions that explicitly incorporate temporal information remains an interesting direction for future work.}

\lucas{A limitation of the present study is that the validation is based exclusively on synthetic wildfire scenarios. While this approach enables a controlled evaluation of parameter recovery under known conditions, it does not directly assess predictive performance in real wildfire events. In real settings, fire propagation is influenced by time-varying environmental factors, particularly wind, which can change significantly over the course of several days. In contrast, the current approach assumes a stationary wind field, representing a simplification that may limit its applicability to real-world scenarios. Addressing these limitations will require incorporating time-dependent environmental inputs and validating the model against observed wildfire data.}

\lucas{From a practical standpoint, these findings indicate that incorporating additional sources of information, such as temporal fire progression (e.g., isochrones) or time-varying environmental conditions, may be necessary to improve parameter estimation and model reliability. This is particularly relevant for decision-making processes in wildfire managment, where accurate predictions of fire spread dynamics are essential for planning and risk assessment.}

\section{Conclusions}

Wildfires are a major concern year after year in the Patagonian region. There is an urgent need to simulate forest fires propagation in real landscapes, taking into account various layers of information—such as wind, slope, and vegetation—in order to test different scenarios for fire prevention. In this work, we calibrated a reaction–diffusion–convection model that simulates fire spread over real maps of the Patagonian region, particularly around Lake Steffen Martin, which is highly representative of the area. We progressively added complexity to the parameter search space in order to evaluate the performance of the fitting method when adjusting the simulations to the synthetically generated data. In the first experiment, we tuned five parameters, seven in the second, and thirteen in the third. To improve model fitting, we performed a bootstrap analysis over the results of the last generation, quantifying the uncertainties of the best individuals. Then we tuned the GA algorithm with machine learning techniques, to improve even more the fitting methodology. As a result, we were able to recover the original parameters with very good accuracy. The best performances were found in the ignition coordinates, followed by the diffusion and advection coefficients. The ignition and extinction rates remained the most challenging parameters. In the third experiment, where five fuel types were adjusted, only for shrubland and pasture these coefficients were accurately estimated. 

Our systematic approach facilitates the estimation of field parameters in heterogeneous landscapes that are otherwise difficult to quantify. Though applied here to an RDC forest-fire model, the method is applicable in principle to a broad range of models based on cellular automata or partial differential equations. \lucas{These results also highlight the importance of incorporating ecologically relevant information, such as temporal fire dynamics, to improve the interpretability and applicability of wildfire spread models in real-world scenarios.}

The next step will involve fitting the model to real wildfire scars, particularly the Lake Martin fire, whose isochrones will enable the calibration of quantities such as the fire spread rate, diffusion coefficient, and the constants related to wind and slope. \lucas{This extension will also require incorporating time-varying environmental conditions, especially wind, to better represent real fire dynamics}. This work represents an important step toward predicting fire propagation in the Patagonian region. Combined with the already developed visual interface, it will serve as a valuable tool for planning and prevention efforts by firefighters and national fire management authorities.

\section{Acknowledgments}

M. D., K. L., and A. K. are members of CONICET. M. D. and K. L. are part of the project grants PI UNRN 40-B-745 and Proyecto Federal de Innovacion: Proyecto RN-2- PFI 2022. L.B. has a scholarship from Instituto Balseiro-UNCUYO-CNEA. We thank Marcelo Bari, Paula Presti and members of ICE for fruitful discussions. Computational resources were provided by the HPC cluster of the Physics Department at Centro Atómico Bariloche (CNEA) and UNC Supercómputo (CCAD), which is part of SNCAD, Argentina. We thank the reviewers for their valuable feedback.

\clearpage
\appendix
\section{Numerical Implementation}\label{appendix:numerics}

The explicit scheme is implemented as follows:

\begin{align}
      S^{n+1/2}_{i,j} &= S^{n}_{i,j} - \beta_{i,j} S^{n}_{i,j} I^{n}_{i,j} \frac{\Delta t}{2} \\
      I^{n+1/2}_{i,j} &= I^{n}_{i,j} + \left( \beta_{i,j} S^{n}_{i,j} I^{n}_{i,j} - \gamma_{i,j} I^{n}_{i,j} \right) \frac{\Delta t}{2} - \vec{v} \cdot \nabla I^{n}_{i,j} \frac{\Delta t}{2} \\
      R^{n+1/2}_{i,j} &= R^{n}_{i,j} + \gamma_{i,j} I^{n}_{i,j} \frac{\Delta t}{2}
\end{align}

\noindent where the gradient operator $\nabla I^{n}_{i,j} = \left(\frac{\partial I^{n}_{i,j}}{\partial x}, \frac{\partial I^{n}_{i,j}}{\partial y} \right)$ is discretized using an upwind method:

\begin{equation}
      \frac{\partial I^{n}_{i,j}}{\partial x} = \frac{1}{\Delta}
            \begin{cases}
                  I^{n}_{i,j} - I^{n}_{i-1,j}, & \text{if } v^{n}_{x_{i,j}} > 0 \\
                  I^{n}_{i+1,j} - I^{n}_{i,j}, & \text{if } v^{n}_{x_{i,j}} \leq 0
            \end{cases}
\end{equation}

\begin{equation}
      \frac{\partial I^{n}_{i,j}}{\partial y} = \frac{1}{\Delta}
            \begin{cases}
                  I^{n}_{i,j} - I^{n}_{i,j-1}, & \text{if } v^{n}_{y_{i,j}} > 0 \\
                  I^{n}_{i,j+1} - I^{n}_{i,j}, & \text{if } v^{n}_{y_{i,j}} \leq 0 
            \end{cases}
\end{equation}

\noindent with $\Delta = \Delta x = \Delta y$. The ADI scheme is implemented as follows:

\begin{align}
      I^{n+1/2}_{i,j} &= I^{n}_{i,j} + \frac{1}{2} \alpha \left( \delta^2_x I^{n+1/2}_{i,j} + \delta^2_y I^{n}_{i,j} \right) \label{eq:adi_first_half}\\
      I^{n+1}_{i,j} &= I^{n+1/2}_{i,j} + \frac{1}{2} \alpha \left( \delta^2_x I^{n+1/2}_{i,j} + \delta^2_y I^{n+1}_{i,j} \right) \label{eq:adi_second_half}
\end{align}

\noindent where $\alpha \equiv \frac{D \Delta t}{\Delta^2}$ and $\delta_x$ and $\delta_y$ are the discretized diffusion operators in $x$ and $y$ directions, respectively.

\begin{align}
      \delta^2_x I^{n}_{i,j} &= I^{n}_{i+1,j} - 2I^{n}_{i,j} + I^{n}_{i-1,j} \\
      \delta^2_y I^{n}_{i,j} &= I^{n}_{i,j+1} - 2I^{n}_{i,j} + I^{n}_{i,j-1}
\end{align}

The resulting matrix equations \ref{eq:adi_first_half} and \ref{eq:adi_second_half} have the advantage to be tridiagonal, so they can be solved with a Thomas algorithm \citep{press2007numerical} for example, that is highly parallelizable and fast in GPU, differing from the Crank-Nicolson method in which the matrix is pentadiagonal and more difficult to solve. The ADI scheme also has the advantages of being second order accurate in both time and space.

The RDC model was implemented on GPUs using the CuPy library \citep{nishino2017cupy}. This implementation enables parallelization of the simulations and significantly reduces execution time. In the code, each grid cell is updated at every time step by a CUDA kernel running a large number of parallel threads. Moreover, to support the genetic algorithm, the program can update multiple maps simultaneously in batch mode.

\section{Genetic Algorithm Implementation}\label{appendix:ga} 

The GA we implemented was a real-coded GA \citep{katoch2021review}. The algorithm operates on a population of candidate solutions, represented as real-valued vectors. Each individual in the population corresponds to a set of parameters for the fire propagation model. The fitness of each candidate is evaluated using Equation \ref{eq:fitness_function}. 

After calculating the fitness for each individual, we applied three genetic operators: selection, crossover, and mutation. 

The selection operator aims to choose the most promising candidates (parents) to generate offspring. We used tournament selection, where a group of $N$ candidates (three in our case) is randomly sampled, and the best one (with minimum fitness) is selected as a parent.  

The crossover operator combines the genetic information of two parents to generate new offspring. We implemented a one-point crossover: a random crossover point is chosen, and the genes are exchanged between the two parents to create two new individuals.  

The mutation operator introduces small random changes to an individual's genetic material to maintain diversity in the population. We applied a Gaussian mutation, in which a random value drawn from a Gaussian distribution is added to a gene with a given probability.

\clearpage

\bibliographystyle{elsarticle-harv} 
\bibliography{references}

@article{barbera2025biotic,
  title={Biotic and physical drivers of fire in northwestern Patagonia},
  author={Barber{\'a}, Iv{\'a}n and Cingolani, Ana Mar{\'\i}a and Tiribelli, Florencia and Mermoz, M{\'o}nica Alicia and Morales, Juan Manuel and Kitzberger, Thomas},
  journal={Fire Ecology},
  volume={21},
  number={1},
  pages={1--21},
  year={2025},
  publisher={Springer}
}

@techreport{bari2022SteffenMartin,
  author       = {Marcelo Bari and Paula Presti and Anabella Carp and Mariana Lipori},
  title        = {Técnicas para el abordaje de la evaluación y predicción del comportamiento del fuego: Experiencia del equipo conformado para el incendio Steffen--Martin, Parque Nacional Nahuel Huapi, diciembre 2021 -- marzo 2022},
  institution  = {Administración de Parques Nacionales, Argentina},
  year         = {2022},
  address      = {San Carlos de Bariloche, Río Negro},
  number       = {IF-2022-37711237-APN-PNNH\#APNAC},
  note         = {Informe técnico del Parque Nacional Nahuel Huapi, Dirección Regional Patagonia Norte, Dirección Nacional de Conservación},
}

@inproceedings{chen2016xgboost,
  title={Xgboost: A scalable tree boosting system},
  author={Chen, Tianqi and Guestrin, Carlos},
  booktitle={Proceedings of the 22nd acm sigkdd international conference on knowledge discovery and data mining},
  pages={785--794},
  year={2016}
}

@article{costafreda2017human,
  title={Human-caused fire occurrence modelling in perspective: a review},
  author={Costafreda-Aumedes, Sergi and Comas, Carles and Vega-Garcia, Cristina},
  journal={International Journal of Wildland Fire},
  volume={26},
  number={12},
  pages={983--998},
  year={2017},
  publisher={CSIRO Publishing}
}

@phdthesis{denham2009prediccion,
  title        = {Predicci{\'o}n de la evoluci{\'o}n de los incendios forestales guiada din{\'a}micamente por los datos},
  author       = {Denham, M{\'o}nica Mal{\'e}n},
  school       = {Universitat Aut{\`o}noma de Barcelona},
  year         = {2009},
  address      = {Barcelona},
}

@article{denham2018using,
  title={Using efficient parallelization in graphic processing units to parameterize stochastic fire propagation models},
  author={Denham, M{\'o}nica Mal{\'e}n and Laneri, Karina},
  journal={Journal of Computational Science},
  volume={25},
  pages={76--88},
  year={2018},
  publisher={Elsevier}
}

@article{denham2022visualization,
  title={Visualization and modeling of forest fire propagation in Patagonia},
  author={Denham, M{\'o}nica Mal{\'e}n and Waidelich, Sigfrido and Laneri, Karina},
  journal={Environmental Modelling \& Software},
  volume={158},
  pages={105526},
  year={2022},
  publisher={Elsevier}
}

@article{fuhrer2000forest,
  title={Forest functions, ecosystem stability and management},
  author={F{\"u}hrer, Erwin},
  journal={Forest Ecology and management},
  volume={132},
  number={1},
  pages={29--38},
  year={2000},
  publisher={Elsevier}
}

@misc{GobRionegro_Incendio_2025,
  author       = {{Gobierno de Río Negro}},
  title        = {Incendio en Los Manzanos: un trabajo constante en condiciones adversas},
  url          = {https://rionegro.gov.ar/articulo/52573/incendio-en-los-manzanos-un-trabajo-constante-en-condiciones-adversas},
  year         = {2025},
  note         = {Fecha: 21 de enero de 2025}  
}

@article{grinsztajn2022tree,
  title={Why do tree-based models still outperform deep learning on typical tabular data?},
  author={Grinsztajn, L{\'e}o and Oyallon, Edouard and Varoquaux, Ga{\"e}l},
  journal={Advances in neural information processing systems},
  volume={35},
  pages={507--520},
  year={2022}
}

@techreport{hoover2021wildfire,
  title={Wildfire statistics},
  institution={Congressional Research Service},
  author={Hoover, Katie and Hanson, Laura A},
  year={2021}
}

@article{jellouli2022impact,
  title={The impact of dynamic wind flow behavior on forest fire spread using cellular automata: application to the watershed BOUKHALEF (Morocco)},
  author={Jellouli, Omar and Bernoussi, Abdes Samed},
  journal={Ecological Modelling},
  volume={468},
  pages={109938},
  year={2022},
  publisher={Elsevier}
}

@article{katoch2021review,
  title={A review on genetic algorithm: past, present, and future},
  author={Katoch, Sourabh and Chauhan, Sumit Singh and Kumar, Vijay},
  journal={Multimedia tools and applications},
  volume={80},
  number={5},
  pages={8091--8126},
  year={2021},
  publisher={Springer}
}

@article{kelso2015techniques,
  title={Techniques for evaluating wildfire simulators via the simulation of historical fires using the Australis simulator},
  author={Kelso, Joel K and Mellor, Drew and Murphy, Mary E and Milne, George J},
  journal={International Journal of Wildland Fire},
  volume={24},
  number={6},
  pages={784--797},
  year={2015},
  publisher={CSIRO Publishing}
}

@article{kitzberger2022projections,
  title={Projections of fire probability and ecosystem vulnerability under 21st century climate across a trans-Andean productivity gradient in Patagonia},
  author={Kitzberger, Thomas and Tiribelli, Florencia and Barber{\'a}, Iv{\'a}n and Gowda, Juan Haridas and Morales, Juan Manuel and Zalazar, Laura and Paritsis, Juan},
  journal={Science of the total environment},
  volume={839},
  pages={156303},
  year={2022},
  publisher={Elsevier}
}

@article{kitzberger2025novel,
  title={A novel fire regime driven by increased lightning activity and lightning ignition efficiency for northwestern Patagonia, Argentina},
  author={Kitzberger, Thomas and B{\"u}rgesser, Rodrigo E},
  journal={International Journal of Wildland Fire},
  volume={34},
  number={9},
  year={2025},
  publisher={CSIRO Publishing}
}

@article{kolaitis2023comparative,
  title={Comparative assessment of wildland fire rate of spread models: effects of wind velocity},
  author={Kolaitis, Dionysios I and Pallikarakis, Christos and Founti, Maria A},
  journal={Fire},
  volume={6},
  number={5},
  pages={188},
  year={2023},
  publisher={MDPI}
}

@article{kolton2019,
   title={Rough infection fronts in a random medium},
   volume={92},
   ISSN={1434-6036},
   url={http://dx.doi.org/10.1140/epjb/e2019-90582-3},
   DOI={10.1140/epjb/e2019-90582-3},
   number={6},
   journal={The European Physical Journal B},
   publisher={Springer Science and Business Media LLC},
   author={Kolton, Alejandro B. and Laneri, Karina},
   year={2019},
   month=jun 
}

@article{laneri2020first,
  title={First steps towards a dynamical model for forest fire behaviour in Argentinian landscapes},
  author={Laneri, Karina Fabiana and Waidelich, Sigfrido and Zimmerman, Viviana Beatriz and Denham, M{\'o}nica Malen},
  journal={Journal of Computer Science \& Technology},
  year={2020},
  publisher={Universidad Nacional de La Plata. Facultad de Inform{\'a}tica}
}

@article{li2024projecting,
  title={Projecting large fires in the western US with an interpretable and accurate hybrid machine learning method},
  author={Li, Fa and Zhu, Qing and Yuan, Kunxiaojia and Ji, Fujiang and Paul, Arindam and Lee, Peng and Radeloff, Volker C and Chen, Min},
  journal={Earth's Future},
  volume={12},
  number={10},
  pages={e2024EF004588},
  year={2024},
  publisher={Wiley Online Library}
}

@article{morales2015stochastic,
  title={A stochastic fire spread model for north Patagonia based on fire occurrence maps},
  author={Morales, Juan Manuel and Mermoz, M{\'o}nica and Gowda, Juan Haridas and Kitzberger, Thomas},
  journal={Ecological Modelling},
  volume={300},
  pages={73--80},
  year={2015},
  publisher={Elsevier}
}

@book{murray2003mathematical,
  author    = {J. D. Murray},
  title     = {Mathematical Biology II: Spatial Models and Biomedical Applications},
  publisher = {Springer},
  series    = {Interdisciplinary Applied Mathematics},
  volume    = {18},
  edition   = {3rd},
  year      = {2003},
  address   = {New York},
  doi       = {10.1007/b98869},
  isbn      = {978-0-387-95228-4}
}

@article{nishino2017cupy,
  title={Cupy: A numpy-compatible library for nvidia gpu calculations},
  author={Nishino, ROYUD and Loomis, Shohei Hido Crissman},
  journal={31st confernce on neural information processing systems},
  volume={151},
  number={7},
  year={2017}
}

@article{shwartz2022tabular,
  title={Tabular data: Deep learning is not all you need},
  author={Shwartz-Ziv, Ravid and Armon, Amitai},
  journal={Information fusion},
  volume={81},
  pages={84--90},
  year={2022},
  publisher={Elsevier}
}

@article{syphard2015location,
  title={Location, timing and extent of wildfire vary by cause of ignition},
  author={Syphard, Alexandra D and Keeley, Jon E},
  journal={International Journal of Wildland Fire},
  volume={24},
  number={1},
  pages={37--47},
  year={2015},
  publisher={CSIRO Publishing}
}

@article{owid-forest-area,
    author = {Hannah Ritchie},
    title = {Forest area},
    journal = {Our World in Data},
    year = {2021},
    url = {https://ourworldindata.org/forest-area},
    note = {Accessed in August 2025}
}

@book{press2007numerical,
  title={Numerical recipes 3rd edition: The art of scientific computing},
  author={Press, William H},
  year={2007},
  publisher={Cambridge university press}
}

@Article{Provatas1995,
author={Provatas, Nikolas
and Ala-Nissila, Tapio
and Grant, Martin
and Elder, K. R.
and Pich{\'e}, Luc},
title={Scaling, propagation, and kinetic roughening of flame fronts in random media},
journal={Journal of Statistical Physics},
year={1995},
month={Nov},
day={01},
volume={81},
number={3},
pages={737-759},
issn={1572-9613},
doi={10.1007/BF02179255},
url={https://doi.org/10.1007/BF02179255}
}

@book{rothermel1972mathematical,
  title={A mathematical model for predicting fire spread in wildland fuels},
  author={Rothermel, Richard C},
  volume={115},
  year={1972},
  publisher={Intermountain Forest \& Range Experiment Station, Forest Service, US}
}

@misc{SNMF2025,
  author       = {{Gobierno Argentino}},
  title        = {Servicio Nacional de Manejo del Fuego, Reporte técnico de ocurrencia},
  year         = {2025},
  url          = {https://www.argentina.gob.ar/seguridad/servicio-nacional-de-manejo-del-fuego/evaluacion-de-peligro-y-alerta-temprana/reporte},
  note         = {Accessed in April 2025}
}

@article{wadhwani2025integrating,
  title={Integrating deep learning with physics-based model for predicting grassfire spread: R. Wadhwani et al.},
  author={Wadhwani, Rahul and Zhang, Xiaoning and Li, Yizhou and Sutherland, Duncan and Moinuddin, Khalid and Huang, Xinyan},
  journal={Journal of Forestry Research},
  volume={36},
  number={1},
  pages={140},
  year={2025},
  publisher={Springer}
}

@misc{WindNinja_USFS_2025,
  title        = {WindNinja: a computer program that computes spatially varying wind fields for wildland fire application},
  author       = {{U.S. Forest Service Research \& Development}},
  url = {https://research.fs.usda.gov/firelab/products/dataandtools/windninja},
  year         = {2025},
  note         = {Release Date June 2 2025; Last updated August 8 2025},
}

@article{yu2025probabilistic,
  title={A probabilistic approach to wildfire spread prediction using a denoising diffusion surrogate model},
  author={Yu, Wenbo and Ghosh, Anirbit and Finn, Tobias Sebastian and Arcucci, Rossella and Bocquet, Marc and Cheng, Sibo},
  journal={arXiv preprint arXiv:2507.00761},
  year={2025}
}

@article{zagarra2024,
   title={Infection fronts in randomly varying transmission-rate media},
   volume={110},
   ISSN={2470-0053},
   url={http://dx.doi.org/10.1103/PhysRevE.110.034308},
   DOI={10.1103/physreve.110.034308},
   number={3},
   journal={Physical Review E},
   publisher={American Physical Society (APS)},
   author={Zagarra, Renzo and Laneri, Karina and Kolton, Alejandro B.},
   year={2024},
   month=sep 
}

\end{document}